\documentclass[epj]{svjour}
\usepackage{graphics}
\begin{document}
\title{Radiative and Collisional Energy Loss, and Photon-Tagged Jets at RHIC}
\author{G.-Y. Qin \and J. Ruppert \and C. Gale \and S. Jeon \and G. D. Moore 
}                     
%
%
\institute{Department of Physics, McGill University, 3600 university, Montreal, QC, H3A 2T8, Canada
}
\date{Received: date / Revised version: date}
%
\abstract{The suppression of single jets at high transverse momenta in a quark-gluon plasma is studied at RHIC
energies, and the
additional information provided by a photon tag is included. The energy loss of hard jets traversing through the
medium is evaluated in the AMY formalism, by consistently taking into account the contributions from radiative events
and from elastic collisions at leading order in the coupling. The strongly-interacting medium in these collisions is
modeled with (3+1)-dimensional ideal relativistic hydrodynamics. Putting these ingredients together with a complete
set of photon-production processes, we present a calculation of the nuclear modification of single jets and
photon-tagged jets at RHIC.
} 
\maketitle
\section{Introduction}\label{intro}

A wealth of experimental data generated at the Relativistic Heavy Ion Collider have shown that high-$p_T$ hadrons in
central A+A collisions are significantly suppressed in comparison with those in binary-scaled p+p collisions
\cite{Adcox:2001jp,Adler:2002xw}. Moreover, the disappearance of back-to-back azimuthal correlations has been observed
in central A+A collisions, in contrast to the strong back-to-back correlations in p+p and peripheral A+A collisions
\cite{Adler:2002tq}. Those results have been attributed to the strong interaction between hard partonic jets produced
in the early stage of the collisions and the hot and dense nuclear medium created in the collisions, and they are
commonly referred to as jet quenching \cite{Gyulassy:1993hr}.

A lot of effort has gone into quantitatively calculating the quenching of jets taking place inside the soft nuclear
matter. Experimentally, jet quenching can be quantitatively described by measurements of various quantities. One of
those is the nuclear modification factor $R_{AA}$ of single-hadron spectra in A+A collisions with respect to that in
p+p interactions scaled by the number of binary collisions. In Ref. \cite{Qin:2007zz} we presented a systematic
calculation of $R_{AA}$ for pions in central and non-central Au+Au collisions at $\sqrt{s} = 200$~AGeV by applying the
Arnold, Moore and Yaffe (AMY) \cite{Arnold:2001ms,Arnold:2001ba,Arnold:2002ja} formalism, where the bulk properties of
the medium created in those collision are described by (3+1)-dimensional relativistic ideal hydrodynamics
\cite{Nonaka:2006yn}, which has been shown to give a good description of bulk properties at RHIC. Here, we present our
calculation of both radiative and collisional energy loss in the same framework and compare the relative contribution
to the nuclear suppression at RHIC \cite{Qin:2007rn,Qin:2008ea}.

Additional information on jet quenching may be obtained by performing correlation measurements, i.e., studying the
production of high-$p_T$ hadrons associated with trigger high-$p_T$ particles. One of the most promising suggested triggers is a
high $p_T$ photon \cite{Wang:1996yh,Wang:1996pe}. Hard scatterings in the early stage of the collisions are an
important source of photons. If we trigger on such photons, the transverse energy of the away-side
associated jet will be determined (note, however, that at next-to-leading order, the kinematics will get contributions from
additional $2\to 3$ processes; we postpone this discussion to a later study.). 
It is noted that high-$p_T$ photons may be produced from other sources in A+A
collisions, such as those involving jet--plasma interaction during jet propagation in the medium and
fragmentation of surviving jets after their passing through the medium. In this work, we present a comprehensive study of
jet--photon correlations by taking into account all relevant sources for high $p_T$ photon production.

\section{Jet energy loss}\label{sec:1}

In our approach, jets (quarks and gluons) evolve in the nuclear medium according to a set of Fokker--Planck
type rate equations for their momentum distributions $P(E,t) = {dN(E,t)}/{dE}$. Here we write
down the generic
form of the rate equations \cite{Jeon:2003gi,Turbide:2005fk},
\begin{eqnarray}\label{FP-eq}
\frac{dP_j(E,t)}{dt} \!&=&\! \! \sum_{ab} \! \int \! d\omega \left[P_a(E+\omega,t) \frac{d\Gamma_{a\to
j}(E+\omega,\omega)}{d\omega dt} \right. \nonumber\\ && \left. - P_j(E,t)\frac{d\Gamma_{j\to b}(E,\omega)}{d\omega
dt}\right], \ \ \ \ \ \
\end{eqnarray}
where ${d\Gamma^{j\to a}(E,\omega)}/{d\omega dt}$ is the transition rate for the partonic process $j\to a$, with $E$
the initial jet energy and $\omega$ the energy lost in the process. The $\omega<0$ part of the integration
incorporates the energy-gain channels. The radiative and collisional parts of the transition rates have been
extensively discussed in \cite{Qin:2007zz,Qin:2007rn}.

The initial jet momentum profiles may be computed from pQCD in the factorized formalism. To obtain the momentum
spectra of the produced high-$p_T$ hadrons in Au+Au collisions at RHIC, the energy loss of partonic jets in the nuclear
medium must be taken into account. This is performed by calculating the medium-modified fragmentation function
$\tilde{D}_{h/j}(z,\vec{r}_\bot, \phi)$ for a single jet,
\begin{eqnarray}
\label{medium_frag_fun} \tilde{D}_{h/j}(z,\vec{r}_\bot, \phi) \!&=&\!\! \sum_{j'} \!\int\! dp_{j'} \frac{z'}{z}
D_{h/j'}(z') P(p_{j'}|p_j,\vec{r}_\bot, \phi), \ \ \
\end{eqnarray}
where $D_{h/j}(z)$ is the vacuum fragmentation function, and $z = p_h / p_{j}$ and $z' = p_h / p_{j'}$ are two momentum
fractions with $p_h$ the hadron momentum and $p_{j}$($p_{j'}$) the initial (final) jet momentum. 
In the above equation, $P(p_{j'}|p_j,\vec{r}_\bot,
\phi)$ is obtained by solving Eq.~(\ref{FP-eq}), representing the probability of obtaining a jet $j'$ with momentum
$p_{j'}$ from a given jet $j$ with momentum $p_j$. It depends on the path taken by the parton and the medium
profile along that path, which in turn depends on the location of the jet origin $r_\bot$ and its propagation angle
$\phi$ with respect to the reaction plane. Therefore, one must convolve the above expression over all transverse
positions and directions to obtain the final hadron spectra. The initial hard parton densities are determined from the
overlap geometry between two nuclei in the transverse plane of the collision zone.

\begin{figure}[htb]
\begin{center}
\resizebox{0.78\linewidth}{!}{%
\includegraphics{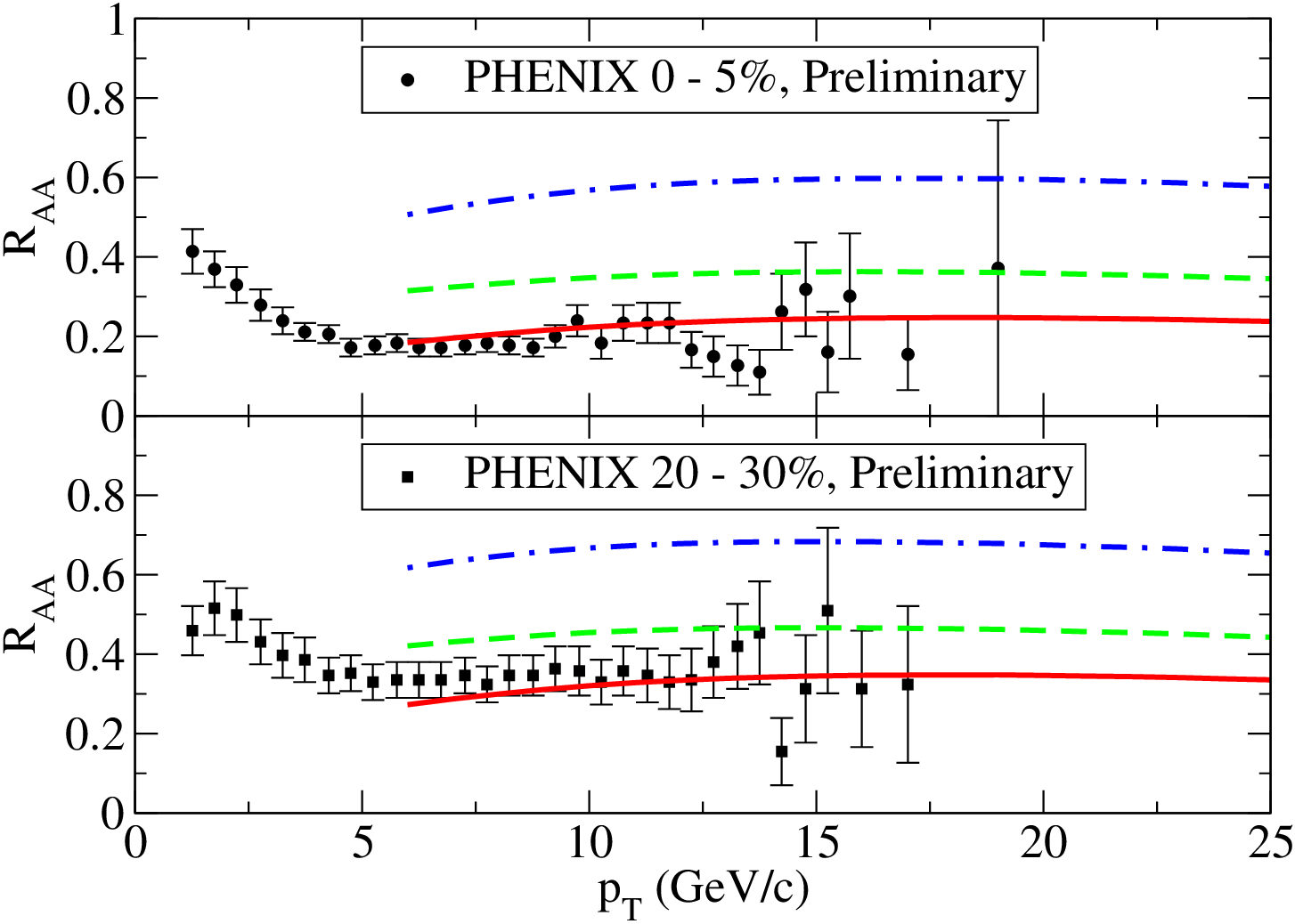}
}\end{center} \caption{The nuclear modification factor $R_{AA}$ for neutral pions in central and mid-central collisions
(from Ref. \cite{Qin:2007rn}). The dashed curves account for only radiative energy loss, the dash-dotted curves for only
collisional energy loss and the solid curves incorporate both radiative and collisional energy loss mechanisms.
 } \label{raa}
\end{figure}

The calculation of $R_{AA}$ for pions measured at mid-rapidity is shown in Fig. \ref{raa} for two different impact
parameters, 2.4 fm and 7.5 fm. The experimental data are taken from PHENIX \cite{Adler:2002xw} for the most central
(0--5\%) and mid-central (20--30\%) collisions. The strong coupling constant $\alpha_s$ is the only
free parameter in our model and it is adjusted in such a way that $R_{AA}$ in most central
collisions (Fig. \ref{raa}, upper panel) is described. The same value, $\alpha_s=0.27$, is used throughout this work for the calculation of photon
production and jet-photon correlations. In this figure, we calculate and compare the relative contribution of induced
gluon radiations and elastic collisions to the final pion $R_{AA}$. One finds that the overall magnitude of $R_{AA}$ is
sensitive to both radiative and collisional energy loss while the shape does not show a strong sensitivity.

\section{Photon production} \label{sec:2}

In nucleus--nucleus collisions, high-$p_T$ photons may be produced from a variety of sources, namely direct
photons, fragmentation photons and jet--plasma photons. Direct photons are produced from hard scatterings in the early
stage of the nuclear collisions, and may be computed from pQCD in the factorized formalism. Fragmentation photons are
produced from the surviving jets after their passing through the soft medium and may be obtained in a
way analogous to the calculation of high-$p_T$ hadrons by replacing the medium-modified hadron fragmentation functions
by appropriated photon fragmentation function.
Jet--plasma photons are those produced from the processes involving the interaction between jets and
the surrounding medium when jets are traversing the plasma. They include induced photon radiation
(bremsstrahlung photons) and the conversion from high energy jets by $2\to 2$ scatterings (conversion photons).
Jet--plasma photons may be
calculated by incorporating an additional evolution equations for photons in Eq.~(\ref{FP-eq}),
\begin{eqnarray}
\label{photon_evolve} \frac{dP^{\rm jet}_\gamma(E,t)}{dt} \!&=&\!\! \int \! d\omega P_{q\bar{q}}(E{+}\omega,t)
\frac{d\Gamma^{\rm jet}_{q\to \gamma}(E{+}\omega,\omega)}{d\omega dt}, \ \ \ \ \ \
\end{eqnarray}
where $d\Gamma^{\rm jet}_{q\to \gamma}(E,\omega)/d\omega dt$ are the transition rates for jet--plasma channels,
including photon bremsstrahlung processes and jet--photon conversion processes, $\Gamma^{\rm jet}_{q\to \gamma} =
\Gamma^{\rm brem}_{q\to \gamma} + \Gamma^{\rm conv}_{q\to \gamma}$. The transition rates for induced photon radiation
have been discussed in Ref. \cite{Arnold:2001ms,Arnold:2001ba,Arnold:2002ja}. The transition rates for jet--photon
conversion processes may be inferred from the photon emission rates in those processes,
\begin{eqnarray}
\frac{d\Gamma^{\rm conv}_{q\to \gamma}(E,\omega)}{d\omega dt} \!&=&\!\! \sum_f \left(\frac{e_f}{e}\right)^2 \frac{2\pi
\alpha_e \alpha_s T^2}{3E} \nonumber\\ && \times \left(\frac{1}{2} \ln \frac{ET}{m_q^2} + C_{2\to 2} \right)
\delta(\omega) . \ \ \
\end{eqnarray}
where $m_q^2 = g_s^2 T^2 / 6$ is the thermal quark mass. In the limit of $E \gg T$, $C_{2\to 2} \approx
-0.36149$ is a constant \cite{Arnold:2002ja}. 
The $\delta$ function generates the constraint that the incoming quarks (or anti-quarks)
experience no energy loss in the conversion processes.

%
\begin{figure}\begin{center}
\resizebox{0.78\linewidth}{!}{%
  \includegraphics{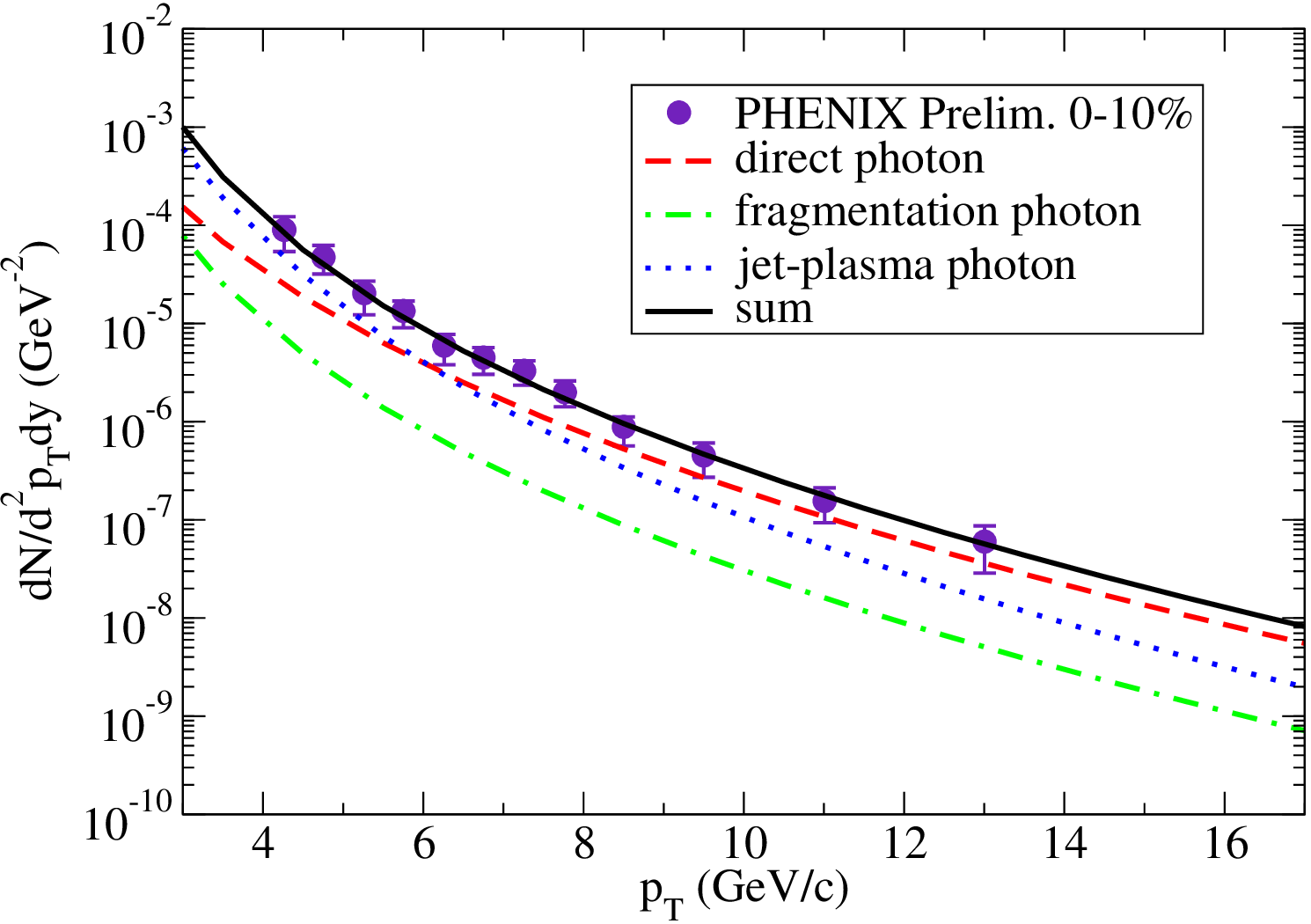}
}\end{center} \caption{  The contributions from different channels to the photon production in Au+Au collisions at
RHIC for $b=2.4$~fm compared with most 0--10\% PHENIX data. }
\label{photon_yield}       
\end{figure}

In Fig. \ref{photon_yield}, the high-$p_T$ photon production from different channels is shown for most central 0--10\%
Au+Au collisions ($b=2.4$~fm) at RHIC, where the number of binary collisions is taken to be $\langle N_{\rm coll}
\rangle = 955$ \cite{Adler:2003qi}. Our calculations are in good agreement with the data taken from PHENIX
\cite{Isobe:2007ku}. While the direct photons dominate at the very high $p_T$ regime ($p_T > 6$~GeV), the presence of the
jet--plasma interaction is important for describing the total photon yield in Au+Au collisions at RHIC, especially 
in the intermediate $p_T$ ($p_T \approx$ 3--6~GeV). The thermal photons are expected to be dominant at very low $p_T$
\cite{Turbide:2005fk,Turbide:2007mi}.

\section{Photon-tagged jets} \label{sec:3}

In correlation studies, one of the most interested observables is the per-trigger yield $P(p_T^{\rm asso}|p_T^{\rm
trig})$,
which represents the conditional probability of producing an associated particle with momentum $p_T^{\rm asso}$ given
a trigger particle with momentum $p_T^{\rm trig}$,
\begin{eqnarray}
P(p_T^{\rm asso}|p_T^{\rm trig}) = \frac{P(p_T^{\rm trig},p_T^{\rm asso})}{P(p_T^{\rm trig})},
\end{eqnarray}
where $P(p_T^{\rm asso})$ and $P(p_T^{\rm trig},p_T^{\rm asso})$ are the single-particle and two-particle joint
probability distributions. As for high-$p_T$ photon--hadron correlations, the associated hadrons are produced from the
fragmentation of survival jets after their passing through the nuclear medium, while the triggered photons may come
from various sources, as has been discussed in the previous section.
Therefore, the photon--hadron correlations depend on jet-photon correlations and the energy loss of the photon-tagged
jets.

To take into account all photon sources, we may write the photon yield as the sum of various parts, each associated
with a specific source,
\begin{eqnarray}
P(p_T^\gamma|p_T^j) = \sum_{\rm src} P(p_T^\gamma, {\rm src}|p_T^j),
\end{eqnarray}
where the sources include direct photons, fragmentation photons and jet--plasma photons. As for direct photons, the
probability distribution is $P(p_T^\gamma, {\rm dir}|p_T^j)  = \delta(p_T^\gamma-p_T^j)$, since
the direct photons have the same momenta as the away-side jets at the production time. The probability function for
fragmentation photons is related to the medium-modified photon fragmentation function by $P(p_T^\gamma, {\rm
frag}|p_T^j)=\tilde{D}_{\gamma/j}(z) /p_T^j$, where $z=p_T^j/p_T^\gamma$ is the momentum fraction. As for jet--plasma
photons, the probability function $P(p_T^\gamma, {\rm jet}|p_T^j)$ may be obtained by solving Eq.~(\ref{photon_evolve})
directly.

\begin{figure}[htb]
\begin{center}
\resizebox{0.78\linewidth}{!}{%
\includegraphics{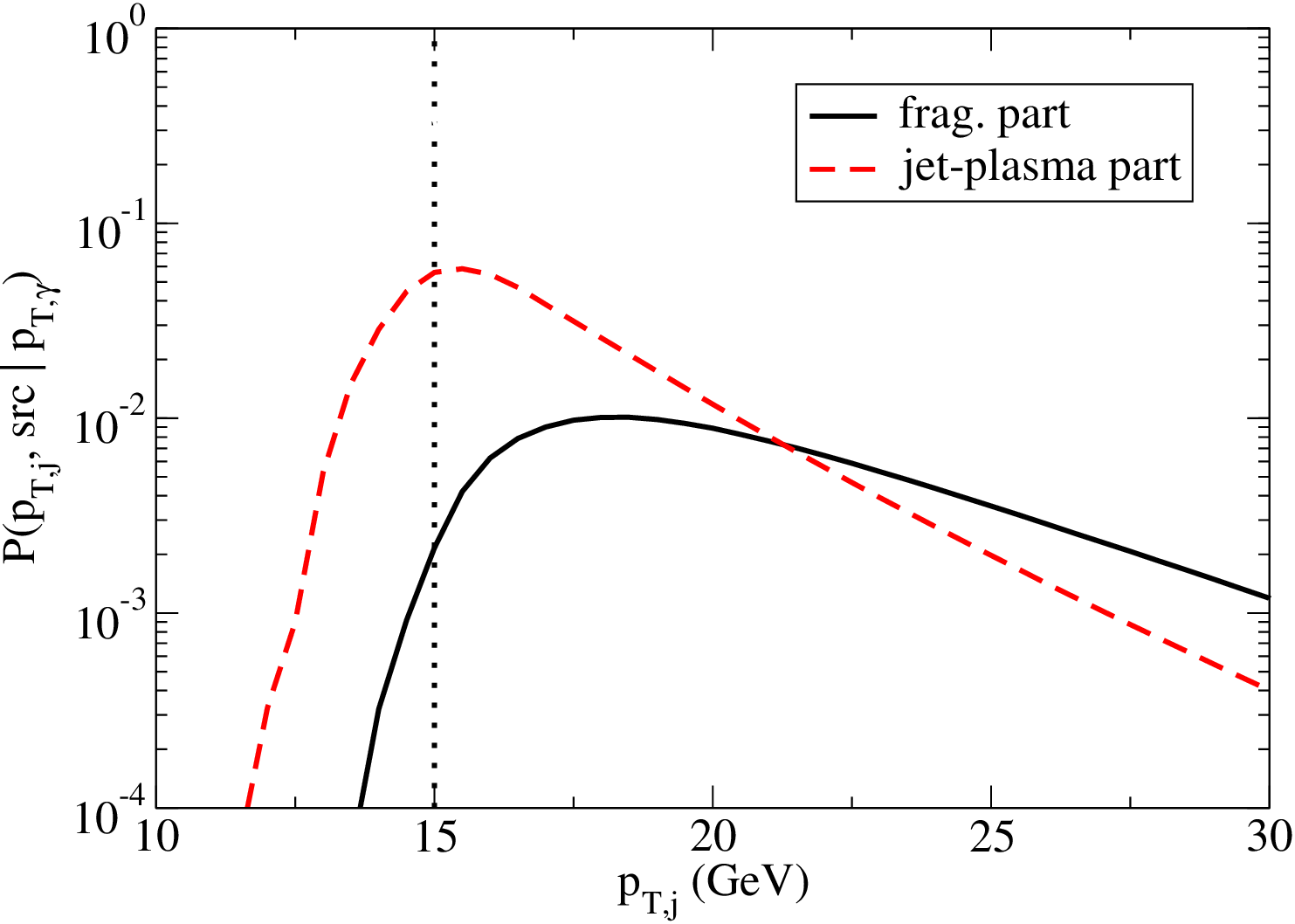}
}\end{center} \caption{The contributions from fragmentation photon and jet--plasma photon parts to the initial jet
momentum distribution at the production time when we trigger on a photon with momentum $p_T^\gamma=15~{\rm GeV}$ in most
central Au+Au collisions at RHIC.} \label{initial_jet_15}
\end{figure}

It would be useful to first take a look at the initial jet momentum profiles at production time, given a trigger photon
with certain transverse momentum. To compare the relative contributions from different photon sources, we may break the
per-trigger yield of the initial jets into different parts,
\begin{eqnarray}
P(p_T^j|p_T^\gamma) = \sum_{\rm src} P(p_T^j, {\rm src}|p_T^\gamma).
\end{eqnarray}
The above probability function $P(p_T^j, {\rm src}|p_T^\gamma)$ is plotted in Fig. \ref{initial_jet_15} for various
photon sources, where the trigger photons have fixed momenta, $p_T^\gamma=15$~GeV. It is obvious that the
probability distribution of jets (quarks plus gluons) tagged by direct photons is a $\delta$ function with some
normalization weight determined by the fraction of direct photons to total photons at $15$~GeV. The momentum profiles
of jets tagged by fragmentation photons and jet--plasma photons are shown by solid and dashed curves.
We find that the initial jets may have higher or lower momenta than the trigger photons. This is attributed to the
energy loss or gain experienced by the jets traversing the thermal medium before producing photons by fragmentation and direct
jet--plasma interaction. It is found that jets tagged by fragmentation photons dominate at high momentum regime, while
jets tagged by jet--plasma photons is prevalent at the relatively lower momentum regime except at $15$~GeV.

\begin{figure}[htb]
\begin{center}
\resizebox{0.78\linewidth}{!}{%
\includegraphics{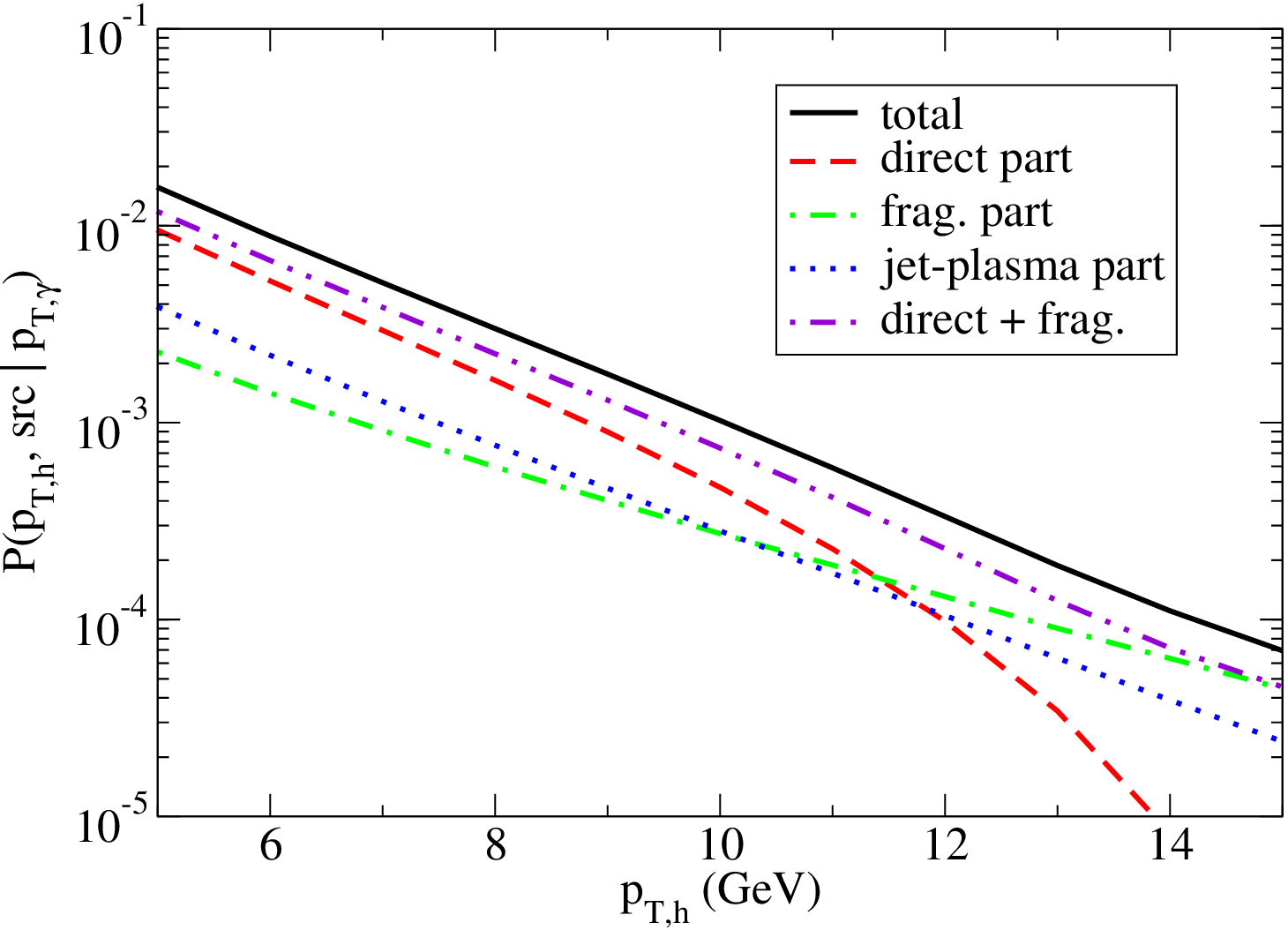}
}\end{center} \caption{Various contributions to per-trigger yield of the away-side hadrons when we trigger a photon
with momentum $p_T^\gamma=15~{\rm GeV}$ in most central Au+Au collisions at RHIC. }
\label{contribution_15}
\end{figure}

By applying the medium-modified fragmentation function in Eq.~(\ref{medium_frag_fun}) to the above jet momentum
profiles, we may obtain the momentum profiles of the associated hadrons. Also, the per-trigger yield of associated
hadrons may be written as the sum of different parts associated with different photon sources,
\begin{eqnarray}
P(p_T^h|p_T^\gamma) = \sum_{\rm src} P(p_T^h, {\rm src}|p_T^\gamma).
\end{eqnarray}
In Fig. \ref{contribution_15}, the above probability function $P(p_T^h, {\rm src}|p_T^\gamma)$ for different sources
of high-$p_T$ photons are plotted as a function of hadron transverse momenta. We observe that the away-side hadrons at
relatively lower $p_T$ regime are mostly produced from those jets that are tagged by direct photons. However, a large amount of
away-side hadrons at the higher $p_T$ regime come from those jets tagged by jet--plasma photons and fragmentation photons.
Especially, close to the trigger photon momentum, the away side hadron production is dominated by those jets tagged by
fragmentation photons. It is noted that the away-side hadrons could have higher momenta than the trigger photon (not
shown).

In the above manuscript, we only present some results for photon-tagged jets in Au+Au collisions at RHIC energies. The
nuclear modification of the photon-triggered hadron production may be obtained by comparing with similar calculations
for p+p collisions. Details of such calculations as well as the comparison to the experimental data will be presented
in an upcoming publication \cite{Qin:2008}.

\section{Conclusions and discussions}

We first calculate the radiative and collisional energy loss of hard partons traversing the
quark--gluon plasma and compare the respective size of these contributions to the nuclear suppression in Au+Au
collisions at RHIC. The evolution of the thermal medium created in those collisions is modeled by utilizing
(3+1)-dimensional relativistic hydrodynamics. It is found that the magnitude of $R_{AA}$ is sensitive to the inclusion
of both radiative and collisional energy loss.

Then we apply the same formalism to study the energy loss of photon-tagged jets in Au+Au collisions at RHIC. For high-$p_T$ photon production, we take into account direct photons as well as fragmentation photons and jet--plasma photons.
Our results illustrate that the interaction between hard partons and the soft medium are significant for the
description of photon data at RHIC. As for the energy loss of photon-tagged jets, we observe the effects
due to jet--plasma interaction and fragmentation. Especially, these sources show a significant contribution to the
photon-hadron correlations at the high-$p_T$ regime for the associated hadrons. Therefore, it is important to include
all photon sources for a full understanding of high-$p_T$ photon--hadron correlations at RHIC. We
conclude that both high transverse momentum jets and photons as well as their correlations are very helpful in
probing the interaction between hard jets and the surrounding hot and dense medium created in
relativistic heavy-ion collisions.
 
\section{Acknowledgements}

We thank C. Nonaka and S. A. Bass for providing their hydrodynamical evolution calculation \cite{Nonaka:2006yn}. 
This work was supported in part by the Natural Sciences and Engineering Research Council of Canada
and by the Fonds Nature et Technologies of Quebec.
 
%
%
\bibliography{qin_reference_list.bib}

\begin{thebibliography}{10}

\bibitem{Adcox:2001jp}
PHENIX, K.~Adcox {\em et~al.},
\newblock Phys. Rev. Lett. {\bf 88}, 022301 (2002), arXiv:nucl-ex/0109003.

\bibitem{Adler:2002xw}
STAR, C.~Adler {\em et~al.},
\newblock Phys. Rev. Lett. {\bf 89}, 202301 (2002), arXiv:nucl-ex/0206011.

\bibitem{Adler:2002tq}
STAR, C.~Adler {\em et~al.},
\newblock Phys. Rev. Lett. {\bf 90}, 082302 (2003), arXiv:nucl-ex/0210033.

\bibitem{Gyulassy:1993hr}
M.~Gyulassy and X.~nian Wang,
\newblock Nucl. Phys. {\bf B420}, 583 (1994), arXiv:nucl-th/9306003.

\bibitem{Qin:2007zz}
G.-Y. Qin {\em et~al.},
\newblock Phys. Rev. {\bf C76}, 064907 (2007), arXiv:0705.2575.

\bibitem{Arnold:2001ms}
P.~Arnold, G.~D. Moore, and L.~G. Yaffe,
\newblock JHEP {\bf 12}, 009 (2001), arXiv:hep-ph/0111107.

\bibitem{Arnold:2001ba}
P.~Arnold, G.~D. Moore, and L.~G. Yaffe,
\newblock JHEP {\bf 11}, 057 (2001), arXiv:hep-ph/0109064.

\bibitem{Arnold:2002ja}
P.~Arnold, G.~D. Moore, and L.~G. Yaffe,
\newblock JHEP {\bf 06}, 030 (2002), arXiv:hep-ph/0204343.

\bibitem{Nonaka:2006yn}
C.~Nonaka and S.~A. Bass,
\newblock Phys. Rev. {\bf C75}, 014902 (2007), arXiv:nucl-th/0607018.

\bibitem{Qin:2007rn}
G.-Y. Qin {\em et~al.},
\newblock Phys. Rev. Lett. {\bf 100}, 072301 (2008), arXiv:0710.0605.

\bibitem{Qin:2008ea}
G.-Y. Qin {\em et~al.},
\newblock (2008), arXiv:0805.4594.

\bibitem{Wang:1996yh}
X.-N. Wang, Z.~Huang, and I.~Sarcevic,
\newblock Phys. Rev. Lett. {\bf 77}, 231 (1996), arXiv:hep-ph/9605213.

\bibitem{Wang:1996pe}
X.-N. Wang and Z.~Huang,
\newblock Phys. Rev. {\bf C55}, 3047 (1997), arXiv:hep-ph/9701227.

\bibitem{Jeon:2003gi}
S.~Jeon and G.~D. Moore,
\newblock Phys. Rev. {\bf C71}, 034901 (2005), arXiv:hep-ph/0309332.

\bibitem{Turbide:2005fk}
S.~Turbide, C.~Gale, S.~Jeon, and G.~D. Moore,
\newblock Phys. Rev. {\bf C72}, 014906 (2005), arXiv:hep-ph/0502248.

\bibitem{Adler:2003qi}
PHENIX, S.~S. Adler {\em et~al.},
\newblock Phys. Rev. Lett. {\bf 91}, 072301 (2003), arXiv:nucl-ex/0304022.

\bibitem{Isobe:2007ku}
PHENIX, T.~Isobe,
\newblock J. Phys. {\bf G34}, S1015 (2007), arXiv:nucl-ex/0701040.

\bibitem{Turbide:2007mi}
S.~Turbide, C.~Gale, E.~Frodermann, and U.~Heinz,
\newblock Phys. Rev. {\bf C77}, 024909 (2008), arXiv:0712.0732.

\bibitem{Qin:2008}
G.-Y. Qin {\em et~al.},
\newblock (in preparation).

\end{thebibliography}
\bibliographystyle{h-physrev5.bst}

%
%

\end{document}